\documentclass[useAMS,usenatbib]{mn2e}

\usepackage{graphicx}

\title[Bipolar ejection by Z And during its 2006 outburst]{Bipolar ejection by the symbiotic binary system Z And during its 2006 outburst\thanks{Based on observations collected at the National Astronomical Observatory Rozhen, Bulgaria}}

\author[N. A. Tomov, M. T. Tomova and D. V. Bisikalo]{N. A. Tomov$^{1}$\thanks{E-mail: tomov@astro.bas.bg }, 
	M. T. Tomova$^{1}$ and
	D. V. Bisikalo$^{2}$\\
$^{1}$Institute of Astronomy, National Astronomical Observatory Rozhen, POBox 136, 4700 Smolyan, Bulgaria\\
$^{2}$Institute of Astronomy of the Russian Academy of Science, 48 Pyatnitskaya Str., 119017 Moscow, Russia
}

\begin{document}
\date{Accepted 2006 December 13. Received 2006 December 12; in original form 2006 November 24}

\pagerange{\pageref{firstpage}--\pageref{lastpage}} \pubyear{2002}

\maketitle

\label{firstpage}

\begin{abstract}
High resolution data in the region of the line H$\alpha$ have been obtained at the time of the light maximum and after it of the 2006 optical outburst of the symbiotic binary Z And. A blue-shifted absorption component indicating outflow velocity of about 1400 km\,s$^{-1}$ as well as additional emission components with similar velocities, situated on the two sides of the main peak of the line were observed during that time. It is suggested that all of them are spectral signature of bipolar outflow, observed for the first time in the optical spectrum of this binary. The emission measure and the mass of the nebular part of the streams were estimated approximately, which reached values of up to about 6\,$\times10^{57}$ cm$^{-3}$ and 8\,$\times10^{-8}$ M$_{\sun}$.
\end{abstract}


\section{Introduction}

Symbiotic stars are interpreted as interacting binaries consisting of a cool visual primary and a compact secondary component accreting matter from the atmosphere of its companion. As a result of accretion the compact object undergoes multiple outbursts accompanied with intensive mass ejection in the form of optically thick shells, stellar wind, single discreet ejections (blobs) as well as collimated bipolar jets. The system Z And is considered as a prototype of the classical symbiotic stars. It consists of a normal cool giant of spectral type M~4.5 \citep{MS}, a hot compact component with a temperature of 1.5$\times10^{5}$ K \citep{Sok06} and an extended circumstellar nebula formed by the winds of the stellar components and partly photoionized by the compact object.

Z And has underwent several active phases (after 1915, 1939, 1960, 1984 and 2000) characterized by repeated optical brightenings, partly realized as a result of a strong redistribution of the energy of the compact object. The energy redistribution was caused by intensive loss of mass \citep*{SS,Boyarchuk,FC95,TTT03,Sok06,Sk06}.
The last active phase of Z And began in the end of August 2000 and continues up to that time. The system underwent four consecutive optical brightenings, whose maxima were in December 2000 ($V\sim8.8$ mag), November 2002 ($V\sim9.8$ mag), September 2004 ($V\sim9.1$ mag) and July 2006 ($V\sim8.6$ mag) \citep[][our $UBV$ data]{SP}. Spectral indications for intensive loss of mass were obtained during three of them, whose maxima were in 2000, 2002 and 2006. The UV line \mbox{P\,{\sc v}} 1117 had a variable P~Cyg profile containing one or two absorption components indicating velocities of 0--300 km\,s$^{-1}$ in November and December 2000. In all cases, however, the absorption reached its maximal depth in the velocity interval 0--50 km\,s$^{-1}$ \citep{Sok06}. During the same time the profiles of the lines \mbox{He\,{\sc i}} were of type P~Cyg too and the absorption component indicated moderate velocities of the flow of 50--100 km\,s$^{-1}$ \citep*{TTZ03,Sk06}. During the period September -- November 2002 the line \mbox{He\,{\sc ii}} 4686, from its side, had two component emission profile consisting of a central narrow component and a broad component indicating stellar wind with a velocity of 1100--1200 km\,s$^{-1}$ \citep*{TTT05}. \citet{Sk06} concluded about intensive loss of mass by the compact object, realized at that time too, analyzing the wings of the line H$\alpha$.
 
Our data show that the lines of \mbox{H\,{\sc i}} and \mbox{He\,{\sc i}} had P~Cyg profile with absorption component indicating moderate velocity during the period July -- September 2006. In some cases the components were more than one. In July, however, these lines had additional absorption component with a high velocity of about 1400 km\,s$^{-1}$ as well. Moreover, during the period July --  September the H$\alpha$ line had additional emission components, placed on the two sides of its central peak and having velocities of 1200--1500 km\,s$^{-1}$. The appearance of the H$\beta$ line was similar. The additional emission line components of Z~And were already reported by \citet{SP} and \citet{SW}, which consider them as indications of bipolar jets. We suppose that all components indicating high velocities appear in bipolar outflow from the outbursting hot object. The 2006 outburst of Z~And is a first one, when optical bipolar outflow from this system was observed. Our note is devoted to this phenomenon, where preliminary results of one study of the H$\alpha$ line are treated.

\section[]{Observations and reduction}

The region of the line H$\alpha$ was observed on eight nights in July, August and September 2006 with the Photometrics CCD camera mounted on the Coude spectrograph of the 2m RCC telescope of the National Astronomical Observatory Rozhen. During the last of these nights besides the region of the line H$\alpha$, that of H$\beta$ was also observed. The spectral resolution was 0.2 \AA\,px$^{-1}$ on all occasions. Some of the exposures were comparatively long to obtain better ratio signal to noise of the continuum although the central emission component of the H$\alpha$ line was saturated in some cases (Table 1). Ever when we made more than one exposure per night, the spectra were added with the aim to improve the signal to noise ratio. The {\sc IRAF} package was used for data reduction. The additional emission components were fitted with a Gaussian to measure their radial velocity and equivalent width. It was made only for those spectra where they are intensive enough (Table 2). The inner uncertainty of the equivalent width in these cases is not more than 50 per cent. The H$\alpha$ flux was obtained using the equivalent width and the $R_{\rmn c}$ flux supposing that it is practically equal to the continuum flux at the position of this line. For this aim we used the $R_{\rmn c}$ magnitudes of \citet{SP} and \citet{SW} taken very close to the time of our observations. The uncertainty
of the continuum flux is not more than 10 per cent and that of the line flux -- not more than 50 per cent. The fluxes were corrected for an interstellar extinction of $E(B-V) = 0.30$ according to the approach of \citet*{Cardelli}. It was used the ephemeris $\rm {Min(vis)=JD}~2\,442\,666^{\rm d}+758\fd8 \times E$, where the orbital period is based on both photometric and spectral data \citep{FL,MK96,Fekel}. 

\begin{table}
 \caption{Journal of observations.}
  \label{journal}
 \begin{tabular}{@{}lccl}
  \hline
	Date & JD$-$ & Orb. & Exp. Time \\
	     & 2\,453\,000 & phase & (min) \\
  
    \hline
  Jul 8 & 924.544 & 0.837 & 15+5 \\
  Jul 9 & 926.431 & 0.840 & 10+20 \\
  Jul 14 & 931.464 & 0.846 & 10+20 \\
  Jul 19 & 935.575 & 0.852 & 10 \\
  Aug 8 & 956.392 & 0.879 & 5+20+10 \\
  Aug 12 & 960.455 & 0.885 & 10+20+20+5 \\
  Sept 7 & 985.594 & 0.918 & 10+20 \\
  Sept 8 & 986.510 & 0.919 & 20 \\
  \hline
 \end{tabular}
\end{table}

\section{Analysis and discussion}

\begin{figure}
\begin{minipage}{8.4cm}
	\resizebox{\hsize}{!}{\includegraphics[]{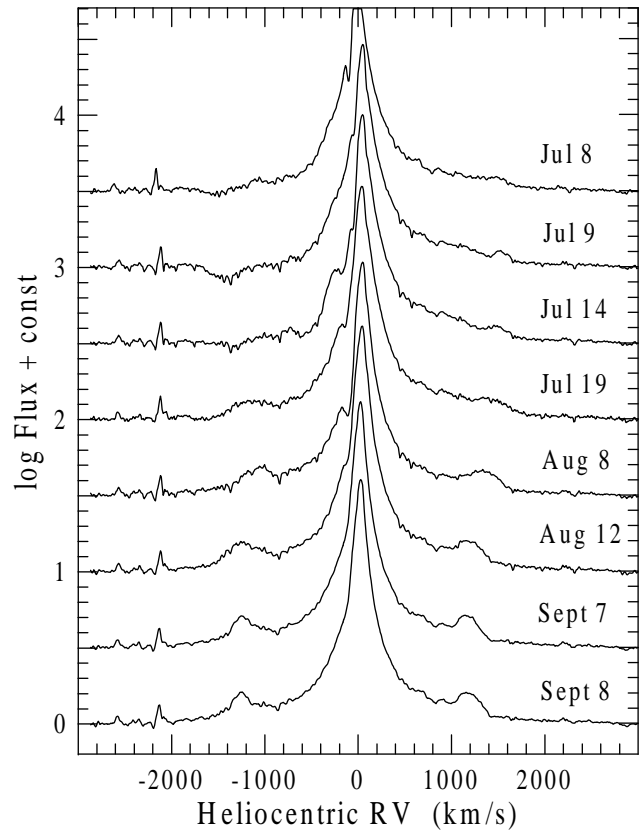}}
    \end{minipage}
%\includegraphics[width=84mm]{plot3c2.ps}
  %% to include a figure, or
 %\vspace{3.5cm}
  %% to leave a blank space
 \caption{Time evolution of the H$\alpha$ line.}
 \label{ha_log}
\end{figure}

In our spectra the line H$\alpha$ had a strong central component located around the reference wavelength, broad component and additional absorption and emission features on both sides of the central component 
(Fig.~\ref{ha_log}). 
The central component was single-peaked having shoulder(s) on its short-wavelengths side, which was not visible only on the spectra taken in September. A weak peak component on the short-wavelengths side of the central component is seen on the spectrum of August 8 and the dip feature between them indicates a moderate velocity of about 100 km\,s$^{-1}$
like the P~Cyg absorption of the He lines. That is why this dip feature can be determined by an outflow too, but
it does not dispose below the level of the continuum probably because of filling by the emission.

The broad component was not located around the reference wavelength during all time of observations and in July its red wing was appreciably more intensive than the blue one. In our treatment, however, we will concentrate on the other additional components of the H$\alpha$ line, which appeared during the period of our observations.

The first spectra in July (Fig.~\ref{ha_log}) show a pronounced absorption with a velocity of $-$1400 km\,s$^{-1}$ on the short-wavelengths side of the central H$\alpha$ component and only a weak emission feature with irregular form and a velocity of about 1500 km\,s$^{-1}$ on its long-wavelengths side. We will consider the absorption first supposing that it is due to mass flow from the compact object in the system. In this case the outflowing material generates an observed photosphere (pseudophotosphere, shell) whose optical depth in the continuum is close to unity. If we assume that the mass flow has spherical symmetry and a constant velocity, this photosphere will be a spherical layer with an internal boundary R equal to the radius of the star and an external boundary of infinity. In this case the density is a function of the distance to the center and is expressed via the continuity equation. The column density $N$ is expressed as 

	\[
	N=\!\!\int_{R}^{\infty}\!\!n(r)\> {\rmn{d}}r = \frac{\dot M}{4\upi \mu m_{\rmn H} \upsilon R}\,\,,
\]
         
\noindent
where $\mu = 1.4$ is a parameter determining the mean molecular weight $\mu m_{\rmn H}$ in the wind \citep{NV87} and the other quantities have their commonly accepted meaning. The optical depth of the layer is $\tau = \kappa N$ , where $\kappa$  is the continuum absorption coefficient (opacity). The mass-loss rate is 

	\[
	\dot M = \frac{4\upi \mu m_{\rmn H} \upsilon R \tau}{\kappa}\,\,.
\]
         
\noindent
Let us calculate the mass-loss rate. The opacity $\kappa$ per one atom is determined from the opacity k$_{\rmn R}$ 
per unit mass by means of the relation $\kappa= {\rmn k}_{\rmn R} \mu m_{\rmn H} $. We used a value of 0.5 cm$^2$\,g$^{-1}$ of the Rosseland mean opacity in the visual region from Table 4 from the work of \citet{Seaton}. In this case $\kappa=1.2\times10^{-24}$ cm$^2$ is obtained. However, we have no estimate of the size of the observed photosphere of the hot component of Z And at the time of our spectral observations. A radius of the hot component of 2.4(d/1.12 kpc) R$_{\sun}$ was obtained in the work of \citet{TTT03} from analysis of the continuum energy distribution of the system at the time of the light maximum of its 2000 outburst. As a result of the analysis of the spectral energy distribution too \citet{Sk06} obtained a radius of 3.1(d/1.5kpc) R$_{\sun}$. The correction of the data of \citet{TTT03} for a distance to the system of 1.5 kpc leads to value of 3.2 R$_{\sun}$ and is in very good agreement with the result of \citet{Sk06}. For the need of our approximate calculation we will use the value of 3.5 R$_{\sun}$, as according to our $UBV$ data the 2006 outburst is more luminous than the 2000 one, when the increase of the light was due mostly to the expansion of the hot component \citep{TTT03}. Having all of these quantities, for the mass-loss rate we obtain $\dot M$=1.33$\times10^{-5}$ M$_{\sun}$\,yr$^{-1}$. 

\begin{figure}
\begin{minipage}{8.4cm}
	\resizebox{\hsize}{!}{\includegraphics[]{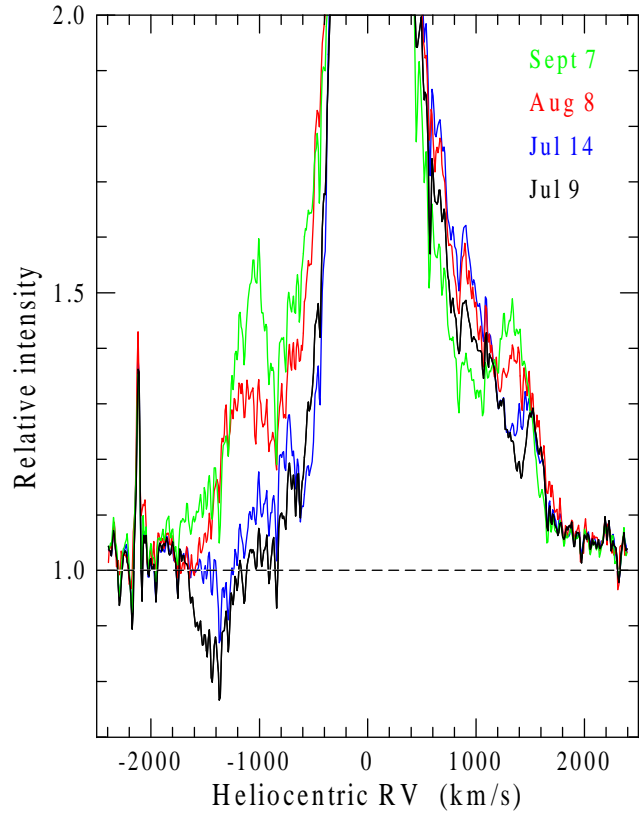}}
    \end{minipage}
 \caption{Transition from the absorption feature to emission one indicating the front stream. The level of the local 
 continuum is marked with a dashed line.}
 \label{abs-em}
\end{figure}

\begin{table*}
 \centering
 \begin{minipage}{150mm}
  \caption{Radial velocities, equivalent widths, line fluxes and emission measures
  					of the outflow components of the H$\alpha$ line.}
  \begin{tabular}{@{}lcccccccc@{}}
  \hline
      & \multicolumn{4}{c}{Blue} & \multicolumn{4}{c}{Red} \\
    Date  & RV & EW & F$\times10^{-12}$ & $n_{\rmn e}^2V\times10^{57}$ 
          & RV & EW & F$\times10^{-12}$ & $n_{\rmn e}^2V\times10^{57}$  \\
        & (km\,s$^{-1}$) & (\AA) & (erg\,cm$^{-2}$\,s$^{-1}$) & (cm$^{-3}$) 
        & (km\,s$^{-1}$) & (\AA) & (erg\,cm$^{-2}$\,s$^{-1}$) & (cm$^{-3}$) \\
 
 \hline
 Jul 19 & $-$1196 & 0.8 & 1.754 & 2.4 & 1445 & 0.9 & 1.813 & 2.5 \\
 Aug 8  & $-$1087 & 1.5 & 2.912 & 3.9 & 1346 & 1.8 & 3.442 & 4.7 \\
 Aug 12 & $-$1260 & 2.5 & 4.828 & 6.5 & 1201 & 2.0 & 3.846 & 5.2 \\
 Sept 7 & $-$1245 & 1.8 & 3.109 & 4.2 & 1178 & 2.0 & 3.362 & 4.6 \\
 Sept 8 & $-$1262 & 1.7 & 2.881 & 3.9 & 1196 & 2.1 & 3.534 & 4.8 \\
 
\hline
\end{tabular}
\end{minipage}
\end{table*}

This value is too great compared with the observed mass-loss rate of the hot compact component of the symbiotic systems during their active phases, which is most frequently about 10$^{-9}$-- 10$^{-7}$ M$_{\sun}$\,yr$^{-1}$ \citep{VN,AC,GYH}. The great value follows from our assumption for spherical symmetry of the outflowing material and the observed high velocity. It can be in reasonable limits if the material does not flow in all directions, but only in some of them, related to areas with small angular sizes. 
The material can flow mainly in the polar areas, if it is prevented in equatorial one for example by an accretion disk. In this case the velocity in the polar areas increases to provide the mass ejection there. This supposition can be related to the existence of two kinds of absorption components indicating very different expansion velocities. 
We can assume that the dip feature of H$\alpha$ is due to mass flow in the orbital plane with a moderate velocity.
The flowing material, however, is prevented by the accretion disk and the direction of its motion changes. In this way
it is ejected in a direction, perpendicular to the plane of the disk, like the gas flowing in that direction.
All of the material gives rise to the high velocity components of the H$\alpha$ line.
Thus the blue shifted absorption indicates motion probably in the front part of the wind which is projected on the stellar disk. Then we can suppose that the weak emission feature with a velocity of 1500 km\,s$^{-1}$ appears in the nonocculted part of the back component of the wind.

As it is seen from the evolution of the spectrum in Figs.~\ref{ha_log} and \ref{abs-em} the blue-shifted absorption component disappears and an emission rises. Thus two emission components on the two sides of the central peak form in July. The disappearance of the blue-shifted absorption and the rise of an emission is most probably due to decrease of the flow rate and/or increase of the number of the emitting atoms in that part of the wind which does not project on the stellar disk. If the absorption is related to the inner part of the wind and the emission -- to its outer one, not projecting on the stellar disk, the evolution of the spectrum shows that the velocity in the front wind component (stream) decreases with the distance to the star. 

It is seen in Fig.~\ref{ha_log} that the emission on the long-wavelengths side of the central peak increases compared to the continuum. It is seen also in the figure that the velocity of the line decreases from about 1500 km\,s$^{-1}$ to about 1200 km\,s$^{-1}$. Taking into account the fact that the distance covered by the emitting gas increases with time it can be concluded that the velocity in the back stream decreases with the distance to the star too. 

The H$\beta$ line had similar emission components, as it is seen on the unique frame taken in its region in September. 

To know the energy emitted in the H$\alpha$ high velocity components as well as the quantity of the emitting gas we calculated their line flux and emission measure. The emission measure was obtained supposing that the gas medium is optically thin and has a constant density. It is not possible to obtain the electron temperature and the electron density in the region where the outflow components are emitted from observation since we have no indication about the appearance in this region of certain lines giving information for these parameters. We suppose that the electron temperature is 20\,000 K and the mean electron density is about 10$^{-8}$ cm$^{-3}$ in the nebular portion of the streams. Then we used recombination coefficient of 5.956$\times10^{-14}$ cm$^{3}$\,s$^{-1}$ for case B, corresponding to these temperature and density \citep{SH}. We adopted a helium abundance of 0.1 \citep{VN} and a distance to the system of 1.5 kpc according to \citet{MK96}. For the calculation of the emission measure we need to know the state of ionization of helium in the emitting region. The line \mbox{He\,{\sc ii}} 4686 was very weak (as it is seen from our data) during the period of our observations which means that singly ionized helium is dominant in the circumbinary nebula at the times close to the epoch of the maximal light. That is why we assume the state of ionization to be He$^+$ in the region of the streams. 

The line flux and the emission measure are listed in Table 2. These data propose that the ejected matter increases till August 12 and after that decreases, which is probably due to decrease of the flow rate. We calculated also the mass of the nebular portion of the streams on the basis of the greatest emission measure observed on August 12. It turned out to be about 8$\times10^{-8}$ M$_{\sun}$ for the front stream and 6$\times10^{-8}$ M$_{\sun}$ for the back one.

\section{Conclusions}

We present results of high resolution spectral observations carried out in the region of the line H$\alpha$ of the spectrum of the symbiotic binary Z And close to the light maximum of its 2006 optical brightening. The line H$\alpha$ had a blue-shifted absorption component with a velocity of about 1400 km\,s$^{-1}$ at the time of maximal light in July. This line had also additional emission components placed at 1200--1500 km\,s$^{-1}$ on the two sides of its central peak in July, August and September. The line H$\beta$ had similar components as far as it is seen on its unique frame. It is supposed that both the absorption and the emission components are signature of bipolar wind outflow from  the hot compact object in this system. The different velocities observed at different spectra indicate probably the decrease of the flow velocity.
 
The energy flux of the outflow components of the H$\alpha$ line and their emission measure were calculated on the basis of $R_{\rmn c}$ photometric data from the literature taken during the time of the spectral observations. These quantities increased till August 12 and after that decreased, which is probably due to diminution of the flow rate. The mass of the nebular part of the streams on August 12 was obtained, which amounts to 6--8$\times10^{-8}$ M$_{\sun}$.

\section*{Acknowledgments}

The authors thank their colleagues Dr. H. Markov, Dr. D. Kjurkchieva, A. Ivanova, B. Spassov and D. Dimitrov who contributed to obtaining the observational data. They are also grateful to the referee Prof. J. Mikolajewska whose remarks contributed to the improvement of the paper.

\bsp

\label{lastpage}

\end{document}